\documentclass[preprint,eqsecnum,superscriptaddress]{revtex4}
\usepackage[T1]{fontenc}
\usepackage[latin1]{inputenc}
\usepackage{graphicx,amsfonts}
\usepackage{amsmath}
\newcommand{\be}{\begin{equation}}
\newcommand{\ee}{\end{equation}}
\newcommand{\bea}{\begin{eqnarray}}
\newcommand{\eea}{\end{eqnarray}}
\newcommand{\bml}{\begin{subequations}}
\newcommand{\eml}{\end{subequations}}

\newcommand{\vx}{\vec{x}}

\newcommand{\vk}{\vec{k}}

\newcommand{\ep}{\epsilon}

\newcommand{\rf}{{}^{\hbox{\tiny{(4)}}}\tilde{R}}
\newcommand{\rc}{{}^{\hbox{\tiny{(4,1)}}}R}
\newcommand{\rt}{{}^{\hbox{\tiny{(3)}}}R}
\newcommand{\rtt}{{}^{\hbox{\tiny{(3)}}}\tilde{R}}
\newcommand{\tin}[1]{\hbox{\tiny{#1}}}
\newcommand{\gf}{{}^{\hbox{\tiny{(4)}}}\tilde{\Gamma}}
\newcommand{\gt}{{}^{\hbox{\tiny{(3)}}}\Gamma}

\begin{document}
\title{Braneworlds scenarios in a gravity model with higher order spatial three-curvature terms}
\author{F. S. Bemfica}
\email{fabio.bemfica@ect.ufrn.br}
\affiliation{Escola de Ciências e Tecnologia, Universidade Federal
do Rio Grande do Norte Caixa Postal 1524, 59072-970, Natal, Rio
Grande do Norte, Brazil}
\affiliation{Instituto de Física, Universidade de São Paulo
Caixa Postal 66318, 05315-970, São Paulo, SP, Brazil}
\author{M. Dias}
\email{mafd@cern.ch}
\affiliation{Departamento de Ci\^encias Exatas e da Terra, Universidade Federal de S\~ao Paulo,
Diadema, SP, Brazil}
\author{M. Gomes}
\email{mgomes@fma.if.usp.br}
\affiliation{Instituto de Física, Universidade de São Paulo
Caixa Postal 66318, 05315-970, São Paulo, SP, Brazil}
\author{J. M. Hoff da Silva}
\email{hoff@feg.unesp.br}
\affiliation{UNESP - Campus Guaratinguet\'a - DFQ,
Av. Dr. Ariberto Pereira da Cunha, 333 CEP 12516-410, Guaratinguet\'a, SP, Brazil}

\begin{abstract}
In this work we study a Ho\v rava-like five-dimensional model in the context of braneworld theory. The equations of motion of such model are obtained and, within the realm of warped geometry, we show that the model is consistent if and only if $\lambda$ takes its relativistic value $1$. Furthermore, we show that the elimination of problematic terms involving the warp factor second order derivatives are eliminated by imposing detailed balance condition in the bulk. Afterwards, the Israel's junction conditions are computed, allowing the attainment of an effective Lagrangian in the visible brane. In particular, we show that the resultant effective Lagrangian in the brane corresponds to a (3+1)-dimensional Ho\v rava-like model with an emergent positive cosmological constant but without detailed balance condition. Now, restoration of detailed balance condition, at this time imposed over the brane, plays an interesting role by fitting accordingly the sign of the arbitrary constant $\beta$, insuring a positive brane tension and a real energy for the graviton within its dispersion relation. Also, the brane consistency equations are obtained and, as a result, the model admits positive brane tensions in the compactification scheme if, and only if, $\beta$ is negative and detailed balance condition is imposed.
\end{abstract}

\maketitle

\section{Introduction}
\label{sec1}

Modified theories of gravity containing higher order terms in curvature have been proposed in the literature mainly due to the interest in obtaining a renormalizable quantum version of Einstein gravity. We know that the linearized version of general relativity (GR) has a propagator that behaves like
\be
\label{1-1}
\sim \frac{1}{-k^2}\,,
\ee
where $k^2=-{k^{0}}^2+\vk^2$ is the square of the graviton four-momentum while $k^{0}$ and $\vk$ are its energy and linear 3-momentum, respectively. To improve the ultraviolet behavior, a modified model containing terms with higher order products of the curvature has been set. As desired, the modified theory was proven to be renormalizable~\cite{Stelle}. However, the presence of higher order terms in time derivative is known to produce pathologies like ghosts in such models due to the fact that it introduces ``wrong'' sign poles in the propagators. Also, in some situations the theory describes tachyon.

In the context of modified theories of gravity and with the aim of having a unitary model that, moreover, could improve the ultraviolet behavior of the propagator, Ho\v rava proposed a model that basically modifies GR by terms involving only products of a 3-curvature defined on a 3-dimensional space~\cite{Horava2}. In other words, by foliating the (3+1) spacetime (here we will call it $\Sigma$) into 3-dimensional hypersurfaces $\sigma$ of constant time $t$ ($\Sigma\approx\Re\times\sigma$), the action may be schematically written as
\be
\label{1-2}
S_H=\frac{1}{2\kappa^2}\int_{\Re} dt\int_{\sigma}d^3x N\sqrt{q}\left[K^{ij}K_{ij}-\lambda K^2+f(\rt,\rt^{ij},\cdots)\right]\,,
\ee
where $\kappa^2=8\pi G$ and $\lambda$ is an arbitrary parameter that may acquire quantum corrections in the quantum procedure but that must flows to its relativistic value $1$ in the limit of large distances. Also,
\be
\label{1-3}
K_{ij}=\frac{1}{2N}\left(\dot{q}_{ij}-2 D_{(i}N_{j)}\right)
\ee
is the extrinsic curvature, $N$ and $N^i$ are the lapse function and shift vector, respectively, known from the ADM formalism, and $D_i$ is the metric preserving and torsion free covariant derivative compatible with the 3-metric $q_{ij}$ ($q=\det(q_{ij})$). The 3-curvature of $\sigma$ is $\rt$. Beyond $\lambda$, the main modifications in such theory must come out from $f(\rt,\rt^{ij},\cdots)$, what may contain terms like $\rt$, $\rt^2$, $\rt^{ij}\rt_{ij}$, and so on which, in view of its higher order spatial derivatives in the metric, shall improve the ultraviolet behavior of the propagator. In the original model, Ho\v rava imposed what he called detailed balance condition on $f(\rt,\rt^{ij},\cdots)$ to avoid the spreading of coupling constants.\cite{footnote01} However, it was shown that it is an important ingredient in proving renormalizability of the theory~\cite{Orlando}. In fact, the propagator contains a bad ultraviolet behaving term that spoils renormalizability unless a detailed balance condition is imposed on the quadratic terms in curvature $\rt^2$ and $\rt^{ij}\rt_{ij}$~\cite{Bemfica6,Bemfica7}.

In this work, we shall investigate a (Ho\v rava-type) braneworld scenario in the context of a Ho\v rava-like five-dimensional model within warped spacetimes. It is well known that the study of braneworld models in such geometry leads to very important results as the solution of the hierarchy problem~\cite{RS}.  As we shall see, the simple exigency of warped geometry enforces the relativistic choice for the $\lambda$ parameter of Eq. (\ref{1}). Moreover, we are particularly interested in the possibility of setting up only branes endowed with positive tensions. It is important to remark, for example, that at least for isotropic branes, its tension may be understood as its vacuum energy. Hence, a negative brane tension (as appearing in the first Randall-Sundrum model~\cite{RS}) is certainly a bad feature. Several possible ways to escape from the necessity of a negative brane tension have appeared. For instance, in a five-dimensional bulk this requirement is relaxed in scalar-tensorial theories~\cite{AGJ}, as well as in $f(R)$ gravitational models~\cite{ehNOIS}. Here, we show that the presence of Ho\v rava terms in the bulk action is also responsible for the existence of only positive tension branes if the $\beta$ parameter of Eq. (\ref{1}) has a suitable sign, quite adequate with our other results. The tension sign are studied via the so-called braneworld sum rules~\cite{GKL,FR}, a one parameter family of consistency conditions giving necessary constraints to be respected by the braneworld scenario  to construct a well defined model from the gravitational point of view. The extra, transverse, dimension is regarded as a $S^{1}/Z_{2}$ orbifold, consistent with the sum rules program. Finally, we assume, somewhat naively, that the distance between the brane is already fixed. However, in our developments it is always open the possibility of inserting additional scalar fields in the bulk in order to implement a stabilization mechanism~\cite{GW}.

This paper is structured as follows: in the next section we present a (4+1)-dimensional Ho\v rava-like gravity. The equations of motion for such model are developed in section \ref{sec3} and, in the realm of warped geometries, where a position dependent warp factor ensures the spacetime is not trivially separable, we show that this condition is satisfied if only if $\lambda=1$. Within this regime, and since the first derivative of the warp factor is discontinuous, section \ref{sec6} implements the Israel's junction conditions over the branes and demonstrates that the detailed balance condition plays an essencial role regarding a bad behaving term arising from the modified theory. Such junction conditions enable us to calculate, in section \ref{sec7}, an effective Lagrangian for the visible brane. The resulting effective Lagrangian turns out to be a (3+1)-dimensional Ho\v rava-like model with an emerging positive cosmological constant. By exploiting the detailed balance condition again, we show that the theory accepts positive tension if the model arbitrary constant $\beta$ is set negative, the correct sign in view of the graviton dispersion relation. The consistency conditions for the branes are obtained in section \ref{sec8} and, again, brane tension positivity requires $\beta<0$, in accordance with previous results. Summary and conclusions are found in section \ref{sec9}.

\section{Ho\v{r}ava-like model in the bulk}
\label{sec2}

The model we will be working with shall be defined in a five dimensional spacetime, the bulk $\mathcal{M}$ (with installed coordinates labeled by $X^A$ with $A=0,\cdots,4$). As it is usual in the ADM formalism, we may choose a foliation $\mathcal{M}\cong \Re\times\Omega_t$ characterized by a function $t(X^A)=cte$, where the hipersurface $\Omega_t$ is spacelike and its normal vector field $n$ is timelike ($n^An_A=-1$ when the spacetime signature is $-++++$). As originally proposed, Ho\v rava theory shall be defined in the foliation $X^0=t=cte$. In this framework, the model we are interested in is given by the action
\be
\label{1}
S=\frac{1}{2{\kappa^{\hbox{\tiny{(5)}}}}^2}\int_\Re dt\int_{\Omega}d^4x\tilde{N}\sqrt{\tilde{q}}\left(\tilde{K}^{ab}\tilde{K}_{ab}-\lambda\tilde{K}^2+
\rf+\alpha\,\rf^2+\beta\,\rf^{ab}\rf_{ab}\right)\,,
\ee
where
\be
\label{2}
\tilde{K}_{ab}=\frac{1}{2\tilde{N}}\left(\dot{\tilde{q}}_{ab}-2\tilde{D}_{(a}\tilde{N}_{b)}\right)
\ee
is the extrinsic curvature, ${\kappa^{\hbox{\tiny{(5)}}}}^2=8\pi G_5$ is the corresponding coupling constant in 5-dimensions, $\tilde{N}$ and $\tilde{N}^a$ ($a=1,\cdots,4$) are the lapse function and the four components of the shift vector, respectively, $\rf_{ab}$ is the Ricci tensor while $\rf$ is the scalar curvature of the foliation $\Omega$ and $\tilde{D}_a$ is the torsion free metric preserving covariant derivative compatible with the spatial metric $\tilde{q}_{ab}$. To fix notation, we are defining, for a generic space, $R^\alpha_{\mu\nu\beta}=\partial_\nu\Gamma_{\mu\beta}^\alpha-\partial_{\mu}\Gamma_{\nu\beta}^\alpha+\cdots$, while $R_{\mu\beta}=R^\alpha_{\mu\alpha\beta}$. Since we are in a five dimensional extension of Ho\v rava-like gravity, the arbitrary parameter $\lambda$ introduced by Ho\v rava has been kept initially, despite the problematic behavior introduced by its presence in the four dimensional case~\cite{Bemfica7,Henneaux,Blas}.
The metric of the (4+1)-dimensional bulk $\mathcal{M}$ is denoted by $G_{AB}$ and may be decomposed as
\be
\label{3}
G_{AB}=\left(\begin{array}{ccc}-\tilde{N}^2+\tilde{N}^a\tilde{N}_a&&
\tilde{N}_a\\
\\
\tilde{N}_a&&\tilde{q}_{ab}\end{array}\right)\,,
\ee
with inverse
\be
\label{4}
G^{AB}=\left(\begin{array}{rcc}-\frac{1}{\tilde{N}^2}&&
\frac{\tilde{N}^a}{\tilde{N}^2}\\
\\
\frac{\tilde{N}^a}{\tilde{N}^2}&&\tilde{q}^{ab}-\frac{\tilde{N}^a\tilde{N}^b}{\tilde{N}^2}\end{array}\right)\,.
\ee
It will prove interesting to define the first fundamental form
\be
\label{5}
\tilde{q}_{AB}=G_{AB}+n_An_B\,,
\ee
where $n_A=(-\tilde{N},0,0,0,0)$ because $n^A=(1/\tilde{N},-\tilde{N}^a/\tilde{N})$. This enables one to obtain $\tilde{q}_{00}=\tilde{N}^a\tilde{N}_a$, $\tilde{q}_{0a}=\tilde{N}_a$ and $\tilde{q}_{ab}=G_{ab}$ while $\tilde{q}^{0A}=0$ and $\tilde{q}^{ab}\tilde{q}_{bc}=\delta^a_c$.

The above considerations together with the identity $\tilde{N}\sqrt{\tilde{q}}=\sqrt{-G}$, where $\tilde{q}=\det(\tilde{q}_{ab})$ and $G=\det(G_{AB})$, enable one to rewrite the action (\ref{1}), up to a surface term, as
\be
\label{6}
S=\frac{1}{2{\kappa^{\hbox{\tiny{(5)}}}}^2}\int_\Re dt\int_{\Omega}d^4x\sqrt{-G}\left[\rc+(1-\lambda)\tilde{K}^2+
\rf+\alpha\,\rf^2+\beta\,\rf^{ab}\rf_{ab}\right]\, ,
\ee
where the scalar curvature of the (4+1)-dimensional bulk has been labeled by $\rc$.

\section{The equations of motion}
\label{sec3}

We shall now proceed with the variation of the action given in (\ref{6}) to obtain the equations of motion. The independent variables are the lapse function $\tilde{N}$, the four components of the shift vector $\tilde{N}^a$ together with the 10 components of the 4-dimensional spatial metric $\tilde{q}_{ab}$. It follows that
\bea
\label{7}
\delta S&=&\frac{1}{2{\kappa^{\hbox{\tiny{(5)}}}}^2}\int_\Re dt\int_{\Omega}d^4x\sqrt{-G}\left[
M^{AB}\delta G_{AB}+2(1-\lambda)\tilde{K}\delta \tilde{K}\right.\nonumber\\
&&\left.-2\left(\alpha\,\rf\rf^{ab}+\beta\,\rf^{d(a}\rf^{b)}_d\right)\delta \tilde{q}_{ab}
+f^{ab}\delta \rf_{ab}
\right]\,,
\eea
where
\be
\label{8}
M_{AB}\equiv-\rc_{AB}+\frac{G_{AB}}{2}\left[\rc+(1-\lambda)\tilde{K}^2+\alpha\,\rf^2+\beta\,\rf^{ab}\rf_{ab}\right]\,,
\ee
while
\be
\label{9}
f_{ab}\equiv 2\alpha\tilde{q}_{ab}\,\rf+2\beta\,\rf_{ab}\,.
\ee
The process is not over yet since we did not expressed the variation in terms of the independent variables $\tilde{N}$, $\tilde{N}^a$, and $\tilde{q}_{ab}$. We must work out the quantities $\delta\tilde{K}$ and $\delta\rf_{ab}$. The first quantity shall be developed by using the definition of the second fundamental form, or extrinsic curvature, given by~\cite{footnote02}
\be
\label{10}
\tilde{K}_{AB}=\tilde{q}_A^C\tilde{q}_B^D\tilde{\nabla}_C n_D\,\Longrightarrow \tilde{K}=\tilde{\nabla}_A n^{A}\,,
\ee
with $\tilde{\nabla}_A$ being the covariant derivative compatible with the $(4+1)$-metric $G_{AB}$ on $\mathcal{M}$. Notice that $\tilde{q}_A^C$ acts as a projector of any vector field defined on $\mathcal{M}$ onto a vector field defined on $\Omega$ at that point. It is worth saying that in the specific foliation we are working in, where $X^0=t$, $\tilde{K}=\tilde{q}^{AB}\tilde{K}_{AB}=\tilde{q}^{ab}\tilde{K}_{ab}$ and $\tilde{K}^{AB}\tilde{K}_{AB}=\tilde{K}^{ab}\tilde{K}_{ab}$ since $\tilde{q}^{0A}=0$. By one hand, one may show that
\be
\label{11}
\delta\tilde{K}=\tilde{\nabla}_A\delta n^A+\frac{1}{2}G^{AC}n^B\tilde{\nabla}_B\delta G_{AC}\,,
\ee
while
\be
\label{12}
\delta n^A=-n^B\tilde{q}^{AC}\delta\tilde{q}_{BC}-\frac{n^A}{\tilde{N}}\delta\tilde{N}\,,
\ee
and obtain the result
\bea
\label{13}
\int_\Re dt\int_{\Omega}d^4x\sqrt{-G}\tilde{K}\delta\tilde{K}
&=&\int_\Re dt\int_{\Omega}d^4x\sqrt{-G}\left[\tilde{D}^aK\,\left(\frac{\delta\tilde{N}_a}{\tilde{N}}-\frac{\tilde{N}^b}{\tilde{N}}\delta\tilde{q}_{ab}\right)
+\frac{\tilde{\nabla}_nK}{\tilde{N}}\delta\tilde{N}\right.\nonumber\\
&&\left.-\frac{1}{2}\left(\tilde{K}^2+\tilde{\nabla}_n\tilde{K}\right)G^{AB}\delta G_{AB}\right]\,,
\eea
up to surface terms, with $\tilde{\nabla}_n\equiv n^A\tilde{\nabla}_A$. To obtain an expression only in terms of independent variables one must consider the identity
\be
\label{14}
G^{AB}\delta G_{AB}=\frac{2}{\tilde{N}}\delta\tilde{N}+\tilde{q}^{ab}\delta\tilde{q}_{ab}
\ee
in (\ref{13}). On the other hand, remember that~\cite{Wald}
\be
\label{15}
\delta\,\rf_{ab}=\tilde{q}^{cd}\tilde{D}_c\tilde{D}_{(a}\tilde{q}_{b)d}-\frac{\tilde{q}^{cd}}{2}\tilde{D}_c\tilde{D}_d\delta\tilde{q}_{ab}-\frac{1}{2}\tilde{q}^{cd}\tilde{D}_{(a}\tilde{D}_{b)}\delta\tilde{q}_{cd}\,,
\ee
enabling one to obtain, neglecting surface terms,
\be
\label{16}
\int_{\Omega}d^4x\sqrt{\tilde{q}}\tilde{N}f^{ab}\delta\,\rf_{ab}
=\int_{\Omega}d^4x\sqrt{\tilde{q}}V^{ab}\delta\tilde{g}_{ab}\,.
\ee
Here,
\be
\label{17}
V_{ab}\equiv \tilde{D}_{(a}\tilde{D}^c\left(\tilde{N}f_{b)c}\right)
-\frac{\tilde{q}^{cd}}{2}\tilde{D}_{c}\tilde{D}_d\left(\tilde{N}f_{ab}\right)
-\frac{\tilde{q}_{ab}}{2}\tilde{D}_c\tilde{D}_d\left(\tilde{N}f^{cd}\right)\,.
\ee

Equations (\ref{13}), (\ref{14}), and (\ref{17}), together with the equality
\be
\label{18}
M^{AB}\delta G_{AB}=-2\tilde{N}M^{00}\delta\tilde{N}+2\left(M^{0a}+\tilde{N}^aM^{00}\right)\delta\tilde{N}_a
+\left(M^{ab}-\tilde{N}^a\tilde{N}^bM^{00}\right)\delta\tilde{q}_{ab}\,,
\ee
lead (\ref{7}) to
\bea
\label{19}
\delta S&=&\frac{1}{2{\kappa^{\hbox{\tiny{(5)}}}}^2}\int_\Re dt\int_{\Omega}d^4x\sqrt{-G}\left(
-2\delta\tilde{N}\left[\tilde{N}M^{00}+(1-\lambda)\frac{\tilde{K}^2}{\tilde{N}}\right]\right.\nonumber\\
&&+2\delta\tilde{N}_a\left[M^{0a}+\tilde{N}^aM^{00}+(1-\lambda)\frac{\tilde{D}^a\tilde{K}}{\tilde{N}}\right]\nonumber\\
&&+\delta\tilde{q}_{ab}\left\{M^{ab}-\tilde{N}^a\tilde{N}^bM^{00}
+\frac{V^{ab}}{\tilde{N}}-(1-\lambda)\left[2\frac{\tilde{N}^{(a}\tilde{D}^{b)}\tilde{K}}{\tilde{N}}
+\tilde{q}^{ab}\left(\tilde{K}^2+\tilde{\nabla}_n\tilde{K}\right)\right]\right.\nonumber\\
&&-2\alpha\,\rf\rf^{ab}-2\beta\,\rf^{d(a}\,\rf^{b)}_d\Bigg\}
\Bigg)\,.
\eea
The variation of the matter sector may be given by
\bea
\label{19-1}
\delta S_M&=&\frac{1}{2}\int_\Re dt\int_{\Omega}d^4x\sqrt{-G}S^{AB}\delta G_{AB}\nonumber\\
&=&\frac{1}{2}\int_\Re dt\int_{\Omega}d^4x\sqrt{-G}\left[-2\tilde{N}S^{00}\delta\tilde{N}+2\left(S^{0a}+\tilde{N}^aS^{00}\right)\delta\tilde{N}_a\right.\nonumber\\
&+&\left.\left(S^{ab}-\tilde{N}^a\tilde{N}^bS^{00}\right)\delta\tilde{q}_{ab}\right]\,,
\eea
rendering the equations of motion to the final form
\bml
\label{20}
\bea
&&\tilde{N}M^{00}+(1-\lambda)\frac{\tilde{K}^2}{\tilde{N}}=-{\kappa^{\hbox{\tiny{(5)}}}}^2S^{00}\,,\label{20a}\\
&&M^{0a}+\tilde{N}^aM^{00}+(1-\lambda)\frac{\tilde{D}^a\tilde{K}}{\tilde{N}}=-{\kappa^{\hbox{\tiny{(5)}}}}^2\left(S^{0a}+\tilde{N}^aS^{00}\right)\,,\label{20b}\\
&&M^{ab}-\tilde{N}^a\tilde{N}^bM^{00}+\frac{V^{ab}}{\tilde{N}}-(1-\lambda)
\left[2\frac{\tilde{N}^{(a}\tilde{D}^{b)}\tilde{K}}{\tilde{N}}+\tilde{q}^{ab}\left(\tilde{K}^2+\tilde{\nabla}_n\tilde{K}\right)\right]\nonumber\\
&&-2\alpha\,\rf\rf^{ab}-2\beta\,\rf^{d(a}\,\rf^{b)}_d=-{\kappa^{\hbox{\tiny{(5)}}}}^2\left(S^{ab}-\tilde{N}^a\tilde{N}^bS^{00}\right)\,.\label{20c}
\eea
\eml
The stress-energy tensor $S_{AB}$ shall be chosen in accordance with the model one is interested in. In the present model we shall be concerned with the case without cosmological constant in the bulk. In fact, as we will see, it does not preclude the emergency of a cosmological constant on the brane as an effect of the Ho\v rava-like bulk.

\section*{Warped geometry and its effect on \boldmath $\lambda$}

Warped spaces are characterized by the metric
\be
\label{4-1}
dS^2=w(y)g_{\mu\nu}(t,x^i)dx^\mu dx^\nu+dy^2\,,
\ee
where, from now on, $X^A=(t,x^i,y)$ with $i=1,2,3$ and $w(y)$ is the warping factor that ensures the geometry is not totally separable. Usually, the warp factor is written as $W^2$ to denote its positivity. However, for further simplicity, in this work we shall use $w>0$ instead. From the above equation, one may easily verify that $\tilde{q}_{ij}(t,x^i,y)=w(y)q_{ij}(t,x^l)$, $\tilde{q}_{44}=1$. It is also interesting to define $y$ independent fields $\tilde{N}(t,x^i,y)=\sqrt{w(y)}N(x^i,t)$ and $\tilde{N}^a(t,x^i,y)=w(y)N^a(t,x^i)$. Naturally, $N^4=0$. Notice that $q_{ij}$ ($q^{il}q_{lj}=\delta^i_j$) is the 3-metric of the 3-dimensional space foliation $\sigma_{t,y}$ of constant $t$ and $y$. By considering all that, it is straightforward to obtain
\bml
\label{4-2}
\bea
\rf_{44}&=&\frac{3}{4}\frac{(\partial_y w)^2}{w^2}-\frac{3}{2}\frac{\partial^2_y w}{w}\,,\label{4-2a}\\
\rf_{ij}&=&-\frac{1}{4}q_{ij}\left[\frac{(\partial_y w)^2}{w}+2\partial_y^2 w\right]
+\rt_{ij}\,,\label{4-2b}\\
\rf&=&\frac{\rt}{w}-3\frac{\partial_y^2 w}{w}\,,\label{4-2c}\\
\rc_{44}&=&\frac{(\partial_y w)^2}{w^2}-2\frac{\partial_y^2 w}{w}\,,\label{4-2d}\\
\rc_{\mu\nu}&=&R_{\mu\nu}-\frac{1}{2}g_{\mu\nu}\left[\frac{(\partial_y w)^2}{w}+\partial_y^2 w\right]\,,\label{4-2e}\\
\rc&=&\frac{R}{w}-3\frac{\partial_y^2w}{w}-\frac{(\partial_y w)^2}{w^2}\,,\label{4-2g}\\
\rf_{4i}&=&\rc_{4\mu}=0\label{4-2f}\,,
\eea
\eml
where $R_{\mu\nu}$ ($\rt_{ij}$) is the Ricci tensor defined by the the metric $g_{\mu\nu}$ ($q_{ij}$). Moreover, with such metric the extrinsic curvature turns out to be
\bml
\label{4-3}
\bea
\tilde{K}_{44}&=&0\,,\label{4-3a}\\
\tilde{K}_{4i}&=&-\frac{N_i}{4N}\frac{\partial_y w}{\sqrt{w}}\,,\label{4-3b}\\
\tilde{K}_{ij}&=&\sqrt{w}K_{ij}\,,\label{4-3c}
\eea
\eml
what lead us to $\tilde{K}=K/\sqrt{w}$. It is worth saying that $K_{ij}$ and $K$ are written in (\ref{1-3}) and do not depend on $y$.

One of the important consequences of the above quoting is that in a region of vacuum on the bulk and aware of the identity $M^{4\mu}=N^4=0$, equation (\ref{20b}) for $a=4$ goes into the form
\be
\label{4-4}
(1-\lambda)\partial_y w=0\,.
\ee
Note that even if we consider the introduction of the cosmological constant in te bulk, equation (\ref{20b}) will not be affected when $a=4$ since in the warped space $G^{04}=0$. The utmost information obtained from the above equation is that there is a division of two extreme regimes. On the one hand, one accepts the introduction of an arbitrary parameter $\lambda\ne 1$ and ends up with the fact that $w$ is a constant since $\lambda$ is a parameter that does not depend on position, what leads to a geometrically separable spacetime, loosing the advantages provided by the warped formalism. On the other hand, one chooses the relativistic value $\lambda=1$ and continues with the situation where $w$ is dependent on $y$, what is the more interesting case in view of braneworld formalism.

\section*{The relativistic choice \boldmath $\lambda=1$}

We claim that a $y$ dependent $w$ is the physical situation we are looking for, so that we must set $\lambda=1$. In such case, the equations of motion become
\be
\label{5-1}
\rc_{AB}-\frac{G_{AB}}{2}\rc=\tilde{\Lambda}_{AB}+{\kappa^{\hbox{\tiny{(5)}}}}^2S_{AB}\,,
\ee
where
\bml
\label{5-2}
\bea
\tilde{\Lambda}_{0A}&=&\frac{G_{0A}}{2}\left(\alpha\,\rf^2+\beta\,\rf^{ab}\rf_{ab}\right)\,,\label{5-2a}\\
\tilde{\Lambda}_{ab}&=&\frac{\tilde{q}_{ab}}{2}\left(\alpha\,\rf^2+\beta\,\rf^{ab}\rf_{ab}\right)+\frac{V_{ab}}{\tilde{N}}\nonumber\\
&&-2\alpha\,\rf\rf_{ab}-2\beta\,\rf_{d(a}\,\rf_{b)}^d\,.\label{5-2b}
\eea
\eml
As one can observe, the bulk has induced an energy momentum tensor on the brane which depends on the warping factor $w$ and the curvature of the 4-dimensional space $\Omega$. It is clear that $\tilde{\nabla}_A\tilde{\Lambda}^{AB}=0$. Within this regime, one may also show that
\be
\label{5-3}
R_{\mu\nu}-\frac{g_{\mu\nu}}{2}R=\Lambda_{\mu\nu}+{\kappa^{\hbox{\tiny{(5)}}}}^2S_{\mu\nu}\,,
\ee
what is the conventional Einstein equations of general relativity in (3+1)-dimensions modified by $\Lambda$ that, componentwise, reads
\bml
\label{5-4}
\bea
\Lambda_{0\mu}&=&\Lambda_{\mu0}=\frac{wg_{0\mu}}{2}\left(-3\frac{\partial_y^2 w}{w}
+\alpha\,\rf^2+\beta\,\rf^{ab}\,\rf_{ab}\right)\,,\label{5-4a}\\
\Lambda_{ij}&=&\frac{V_{ij}}{\sqrt{w}N}-\frac{2}{w}\left(\alpha\,\rt\,\rt_{ij}+\beta\rt^l_{(i}\,\rt_{j)l}\right)+\frac{6\alpha}{w}\partial^2_y w\,\rt_{ij}\nonumber\\
&+&\frac{wq_{ij}}{2}\left(-3\frac{\partial_y^2 w}{w}
+\alpha\,\rf^2+\beta\,\rf^{ab}\,\rf_{ab}\right)\,.\label{5-4b}
\eea
\eml
One can also verify that
\bea
\label{5-5}
\alpha\,\rf^2+\beta\,\rf^{ab}\,\rf_{ab}&=&
3\frac{(\partial_y^2 w)^2}{w^2}\left(3\alpha+\beta\right)
-\frac{\rt\partial_y^2 w}{w^2}\left(6\alpha+\beta\right)\nonumber\\
&+&\beta\left[\frac{3}{4}\frac{(\partial_y w)^4}{w^4}-\frac{3}{2}\frac{(\partial_y w)^2\partial_y^2 w}{w^3}-\frac{\rt}{2}\frac{(\partial_y w)^2}{w^3}\right]\nonumber\\
&+&\frac{1}{w^2}\left(\alpha\,\rt^2+\beta\,\rt^{ij}\rt_{ij}\right)\,.
\eea
Again, $\nabla_\mu \Lambda^{\mu\nu}=0$, where now $\nabla_\mu$ is compatible with the $y$ independent foliation metric $g_{\mu\nu}$.

\section{Israel's junction conditions and the detailed balance condition}
\label{sec6}

\begin{figure}
\begin{center}
\includegraphics[scale=.5]{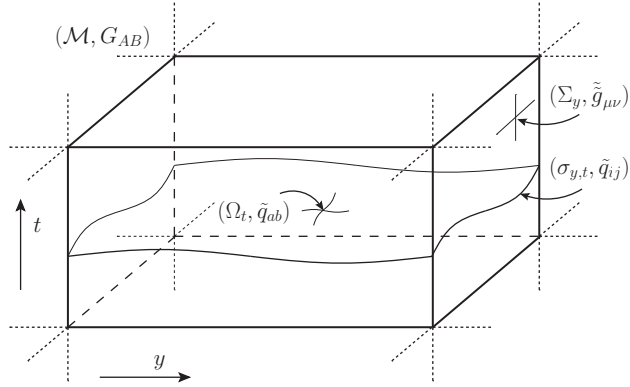}
\caption{\label{Fig1} The whole manifold (five-dimensional bulk) is denoted by $\mathcal{M}$ with metric $G$ and installed coordinates $X^A$, while $\Omega_t$ is the foliation of $\mathcal{M}$ with constant $t$ and metric $\tilde{q}_{ab}$. For constant $y$ we get two such foliations: the one of constante $y$, namely, $\Sigma_t$, with metric $\tilde{\tilde{g}}_{\mu\nu}$ together with $\sigma_{t,y}$, which corresponds to spatial hypersurfaces with $t$ and $y$ simultaneously constant and metric $\tilde{q}_{ij}$.}
\end{center}
\end{figure}

It would be interesting to insert several branes in our scenario, investigating the physical outputs of this type of gravitational theory. In order to do so, let us assume a brane localized, say, at $y=y_r$. In this vein, the spacetime splits in two parts, namely, $\mathcal{M}=\mathcal{M}^{+}_r\cup\mathcal{M}^{-}_r$, with a common boundary $\partial\mathcal{M}^{+}_r\bigcap\partial\mathcal{M}^{-}_r=\Sigma_r$. Due to this separation, we must obtain the metric junction condition in the region $\Sigma$ in each brane. This may be done by integrating Eq.~(\ref{5-3}) from $y_r-\ep$ to $y_r+\ep$ in the limit $\ep\to0$~\cite{Israel}. In fact, the metric must be continuous in this junction. However, because the branes are foliations of the spacetime $\mathcal{M}$ fixed at a constant $y$ whose direction respect $Z_{2}$ symmetry, the vector $\tilde{\tilde{n}}$ ($\tilde{\tilde{n}}^A\tilde{\tilde{n}}_A=1$) normal to the brane must be discontinuous, turning the extrinsic curvature $\tilde{\tilde{K}}_{AB}=(1/2)\mathcal{L}_{\tilde{\tilde{n}}}\tilde{\tilde{g}}_{AB}=(1/2)\left(\tilde{\tilde{n}}^C\partial_C\tilde{\tilde{g}}_{AB}+2\tilde{\tilde{g}}_{C(A}\partial_{B)}\tilde{\tilde{n}}^C\right)$ also discontinuous. In other words,
\be
\label{6-1}
\tilde{\tilde{n}}^{+}=-\tilde{\tilde{n}}^{-}\,,
\ee
and, as a consequence, $\tilde{\tilde{K}}_{AB}^{+}=-\tilde{\tilde{K}}_{AB}^{-}$. In this foliation
\be
\label{61}
\tilde{\tilde{g}}_{AB}\equiv G_{AB}-\tilde{\tilde{n}}_A\tilde{\tilde{n}}_B\,,
\ee
$G_{44}=\tilde{\tilde{N}}^2+\tilde{\tilde{N}}^\mu \tilde{\tilde{N}}_\mu$, $G_{4\mu}=\tilde{\tilde{N}}_\mu$, and $G_{\mu\nu}=\tilde{\tilde{g}}_{\mu\nu}=wg_{\mu\nu}$. By imposing the metric (\ref{4-1}), the components $\mu
,\, \nu$ of the extrinsic curvature can be readily obtained:
\be
\label{6-2}
\tilde{\tilde{K}}_{\mu\nu}=\frac{1}{2\tilde{\tilde{N}}}\left(\partial_y\tilde{g}_{\mu\nu}-2\nabla_{(\mu}\tilde{\tilde{N}}_{\nu)}\right)
=\frac{g_{\mu\nu}}{2}\partial_y w(y)\,.
\ee
Figure \ref{Fig1} illustrates the foliations that so far we dealt with.
The information one gets from what was above mentioned is that although $w(y)$ is continuous throughout $\mathcal{M}$, its first derivative is discontinuous in passing through a brane, namely,
\be
\label{6-3}
\partial_y w^{+}=-\partial_y w^{-}\,.
\ee
Such discontinuity of the warp factor first derivative crossing a brane located at $y_r$ introduces a Dirac delta $\delta(y-y_r)$ in the second derivative of $w$. In particular, the term
\be
\label{6-4}
3\frac{(\partial_y^2 w)^2}{w^2}\left(3\alpha+\beta\right)
\ee
present in Eq.~(\ref{5-5}) will contain a square delta on $y$, sprouting an inconsistency in the model. It is possible to circumvent this problem by imposing  the detailed balance condition~\cite{Horava2} in the quadratic terms in the 4-curvature $\rf$. In four dimensions the Weyl metric can be written as
\be
\label{6-5}
\mathcal{G}^{ab,cd}=\tilde{q}^{a(c}\tilde{q}^{d)b}-\tilde{q}^{ab}\tilde{q}^{cd}
\ee
whose inverse is
\be
\label{6-6}
\mathcal{G}_{ab,cd}=\tilde{q}_{a(c}\tilde{q}_{d)b}-\frac{1}{3}\tilde{q}_{ab}\tilde{q}_{cd}\,.
\ee
For the quadratic terms in the 4-curvature $\rf$ we may impose the  detailed balance condition obtained from the tensor
\be
\label{6-9}
C^{ab}_{4}=\frac{1}{\sqrt{\tilde{q}}}\frac{\delta W^{(4)}}{\delta \tilde{q}_{ab}}=-\rf^{ab}+\frac{\tilde{g}^{ab}}{2}\rf
\ee
which, in turns, arises from the 4-action
\be
\label{6-8}
W^{(4)}=\int_{\Omega}  d^4x \sqrt{\tilde{q}}\rf\,.
\ee
The resulting condition then reads
\be
\label{6-7}
\beta\int dt d^4x\sqrt{G}C^{ab}_{4}\mathcal{G}_{ab,cd}C^{cd}_4=\beta\int dt d^4x\sqrt{G}\left(\rf^{ab}\rf_{ab}-\frac{1}{3}\rf^2\right)\,.
\ee
The relation imposed by (\ref{6-7}) is equivalent of setting
\be
\label{6-10}
\alpha=-\frac{\beta}{3}
\ee
in Eq.~(\ref{1}). From now on we are fixing $\alpha$ as (\ref{6-10}) by imposing detailed balance condition on quadratic terms in the 4-curvature $\rf$. This clearly eliminates the delta squared from (\ref{5-5}).

Let us integrate Eq.~(\ref{5-3}) over $y$ from $y_r-\ep$ to $y_r+\ep$ in the limit $\ep\to0$. For a model with the stress-energy tensor
\be
\label{6-101}
S_{AB}\equiv{T_{\tin{B}}}_{AB}+\sum_{r=0}^{N-1}\delta(y-y_r)\tilde{\tilde{g}}^{\mu}_A\tilde{\tilde{g}}^{\nu}_BS^{(r)}_{\mu\nu}\,,
\ee
with ${T_{\tin{B}}}_{AB}$ being the energy-momentum tensor of the bulk while
\be
\label{6-102}
S^{(r)}_{\mu\nu}=-\tilde{\tilde{g}}_{\mu\nu}\tau^{(r)}+T^{(r)}_{\mu\nu}
\ee
is the localized stress-energy tensor of the $r$-th brane written in term of the brane tension $\tau^{(r)}$ together with its energy-momentum tensor $T^{(r)}_{\mu\nu}$. Continuous terms in $y$ will annulate.
The resulting equations are, then,
\bml
\label{6-11}
\bea
\left(\frac{-3w_r+\beta R}{w_r}\right)\partial_y\ln{w^{+}}-\frac{\beta}{2}(\partial_y \ln{w^{+}})^3&=&-{\kappa^{\hbox{\tiny{(5)}}}}^2\tilde{\tilde{g}}^{\mu0}{S^{\hbox{\tiny{(r)}}}}_{0\mu}=-{\kappa^{\hbox{\tiny{(5)}}}}^2{S^{(r)}}_0^0\,,\label{6-11a}\\
\tilde{\tilde{g}}^{i\mu}{S^{\hbox{\tiny{(r)}}}}_{\mu0}&=&{S^{\hbox{\tiny{(r)}}}}_0^i=0\,,\label{6-11b}\\
\int_{-}^{+}dy\frac{V_{ij}}{\sqrt{w}N}-4\beta\partial_y \ln{w^{+}}\,\rt_{ij}&=&
{\kappa^{\hbox{\tiny{(5)}}}}^2\left(-{S^{\hbox{\tiny{(r)}}}}_{ij}+w_rg_{ij}\,{S^{\hbox{\tiny{(r)}}}}_0^0\right)\,,\label{6-11c}
\eea
\eml
where $w_r$ stands for $w(y_r)$. The equations above not only tell us about the discontinuity of $\partial_y w$ but also impose some conditions on the metric $g_{\mu\nu}$, on the 3-curvature $\rt$, and on the tensor $V_{ij}$. In particular, although ${S^{(r)}}_{0\mu}\ne0$, the combination $\tilde{\tilde{g}}^{i\mu}{S^{(r)}}_{0\mu}$ must vanish.

\section{Effective Lagrangian on the brane}
\label{sec7}

Let us now consider the visible brane, arbitrarily chosen as the $r=0$ brane in the position $y=0$.\cite{footnote03} By first considering Eq.~(\ref{6-11a}), one must find the solution for $A$ by solving
\be
\label{7-2}
\left(\frac{-3w_0+\beta \rt}{w_0}\right)A-\frac{\beta}{2}A^3={\kappa^{\hbox{\tiny{(5)}}}}^2\left(\tau^{\hbox{\tiny{(0)}}}-{T^{\hbox{\tiny{(0)}}}}_0^0\right)\,,
\ee
where
\be
\label{7-3}
A\equiv\partial_y \ln{w^{+}}\,.
\ee
Clearly $\partial_y w^{+}=w_0 A$ and, since we want a decreasing warp factor coming out from the (visible) brane, the next brane must have a smaller warp factor. In other words,
\be
\label{7-4}
A<0\,.
\ee
The equation for $A$ is of third degree and its solution is complicated. Instead of searching such solution, let us analyze possible values for $A$ in regions in our universe where ${T^{\hbox{\tiny{(0)}}}}_0^0=0$. The positivity of the brane tension implies
\be
\label{7-41}
\frac{\beta}{2}|A|^3-\left(\frac{-3w_0+\beta \rt}{w_0}\right)|A|>0\,.
\ee
There are three situations where (\ref{7-41}) can be valid:
\begin{enumerate}
\item $\beta>0$ and $-3w_0+|\beta| \rt>0$. This combination demands
\be
|A|>\sqrt{\frac{2}{w_0|\beta|}(-3w_0+|\beta| \rt)}\,.
\ee
The odd result is that it does not permit solutions as $\rt=0$, what is expected at large distances due to asymptotic flatness;
\item $\beta>0$ and $|\beta| \rt<3w_0$, implying that there are solutions for all
\be
|A|>0\,;
\ee
and
\item $\beta<0$ and $3w_0+|\beta| \rt>0$, consisting in solutions of the type
\be
0<|A|<\sqrt{\frac{2}{w_0|\beta|}(3w_0+|\beta| \rt)}\,.
\ee
\end{enumerate}
One may summarize all three possible solutions by setting
\be
\label{7-5}
A\equiv-\gamma\sqrt{\frac{2\bar{s}}{w_0|\beta|}\left(-s|\beta|\rt+3w_0\right)}\,,
\ee
with $s\equiv \beta/|\beta|=sign(\beta)$ and $\bar{s}=\pm 1$. Clearly, solution 1 relies on the choice $s=1$ and $\bar{s}=-1$ imposing $\gamma>1$, while solution 2 consists of setting $s=1$ and $\bar{s}=1$ for any $\gamma>0$, and, the last possibility, solution 3 corresponds to $s=-1$ and $\bar{s}=1$ and demands $0<\gamma<1$. In terms of the dimensionless $\gamma$, (\ref{7-2}) turns out to be
\be
\label{7-6}
\frac{|\beta|}{2}\left[\frac{2\bar{s}}{w_0|\beta|}\left(-s|\beta|\rt+3w_0\right)\right]^{\frac{3}{2}}
\left(\bar{s}\gamma+s\gamma^3\right)={\kappa^{\hbox{\tiny{(5)}}}}^2\left(\tau^{\hbox{\tiny{(0)}}}-{T^{\hbox{\tiny{(0)}}}}_0^0\right)\,.
\ee
So far we still have a complicated equation in $\gamma$. However, as well as $\gamma$ is an unknown, so is the tension $\tau^{\hbox{\tiny{(0)}}}$ on the brane. Instead of  finding an expression for $\gamma$ as a function of $\tau^{(0)}$, $\rt$, $\beta$, and so forth, we will try to limit some values for $\gamma$ and then express $\tau^{(0)}$ in terms of it. In order to do that we may study the gravity sector of the effective Lagrangian for the brane that, up to surface terms in $y$, is written as follows:
\bea
\label{7-7}
\mathcal{L}_G(t,\vx)&\equiv&\mathcal{L}^{(5)}(t,\vx,y=\pm\ep)\nonumber\\
&=&\frac{N\sqrt{q}}{2{\kappa^{\hbox{\tiny{(5)}}}}^2}\left\{w_0\left(K^{ij}K_{ij}-K^2+
\rt\right)+3(\partial_y w^{\pm})^2-\frac{\beta}{3}\,\rt^2+\beta\,\rt^{ij}\rt_{ij}\right.\nonumber\\
&+&\left.\beta\left[\frac{1}{4}\frac{(\partial_y w^{\pm})^4}{w_0^2}-\frac{\rt}{2}\frac{(\partial_y w^{\pm})^2}{w_0}\right]\right\}\nonumber\\
&=&\frac{N\sqrt{q}}{2{\kappa^{\hbox{\tiny{(5)}}}}^2}\left\{w_0\left[K^{ij}K_{ij}-K^2+
\rt\left(1-9s\bar{s}\gamma^2-6\gamma^4\right)\right]-\beta\rt^2\left(\frac{1}{3}-s\bar{s}\gamma^2-\gamma^4 \right)\right.\nonumber\\    &+&\beta\,\rt^{ij}\rt_{ij}+\frac{9 w_0^2 \gamma ^2}{\beta }\left(2s\bar{s}+\gamma^2\right)\bigg\}\,,
\eea
with $q\equiv\det(q_{ij})$.

At this point a digression is needed. We recall the results found in refs.~\cite{Bemfica6,Bemfica7}, where it was shown that the propagator of the linearized theory given by a Lagrangian like the one above will get an improvement in its ultraviolet behavior if and only if the quadratic terms in the 3-curvature $\rt$ obeys detailed balance condition. Although we have already imposed detailed balance condition in the quadratic term for the Lagrangian in (4+1)-dimensions,  the proceeding of evaluating the effective (3+1)-dimensional Lagrangian has changed it. To restore detailed balance condition, the quadratic term in the 3-curvature must derive from a 3-dimensional action defined in the foliation $\Sigma_0\cong\Re\times\sigma_0$, namely,
\be
\label{7-8}
W^{(3)}=\int_{\sigma_0}d^3x\sqrt{q}\rt\,.
\ee
In three dimensions the inverse of the Weyl metric is given by
\be
\label{7-9}
\mathcal{G}_{ij,kl}=q_{i(k}q_{l)j}-\frac{1}{2}q_{ij}q_{kl}\,.
\ee
Now, the detailed balance condition will come out from the tensor
\be
\label{7-10}
C^{ij}_{3}=\frac{1}{\sqrt{q}}\frac{\delta W^{(3)}}{\delta q_{ij}}=-\rt^{ij}+\frac{q^{ij}}{2}\rt
\ee
and the result is
\be
\label{7-11}
\beta\int_{\Re}dt\int_{\sigma_0}d^3xN\sqrt{q} C^{ij}_3\mathcal{G}_{ij,kl}C^{kl}_3=\beta\int_{\Re}dt\int_{\Sigma_0}d^3x N\sqrt{q}
\left(-\frac{3}{8}\rt^2+\rt^{ij}\rt_{ij}\right)\,.
\ee

A direct consequence of demanding the detailed balance condition in the (3+1) Lagrangian (\ref{7-7}) corresponds to set
\be
\label{7-12}
-\frac{1}{3}+s\bar{s}\gamma^2+\gamma^4=-\frac{3}{8}\,.
\ee
There is no real solution to the above equation when $s=\bar{s}$ and, therefore, solution 2 is not acceptable. For $s=-\bar{s}\Rightarrow s\bar{s}=-1$ (either solution 1 or solution 3), the positive values for $\gamma$ turn out to be
\be
\label{7-13}
\gamma_{\pm}=\frac{1}{2} \sqrt{2\pm\sqrt{\frac{10}{3}}}>0\,.
\ee
Since $0<\gamma_{\pm}<1$, the possible solution that leads to $\tau^{\hbox{\tiny{(0)}}}>0$ is solution 3 and corresponds to $\beta=-|\beta|<0$. In fact,
\be
\label{7-131}
\tau^{\hbox{\tiny{(0)}}}_{\pm}={T^{\hbox{\tiny{(0)}}}}_0^0+\frac{|\beta|}{48{\kappa^{\hbox{\tiny{(5)}}}}^2 w_0} \left(2 \sqrt{3}\mp\sqrt{10}\right) \sqrt{6\pm\sqrt{30}}\left[\frac{2}{w_0|\beta|}\left(|\beta|\rt+3w_0\right)\right]^{\frac{3}{2}}\,.
\ee

The resulting effective Lagrangian then reads
\bea
\label{7-14}
\mathcal{L}_G^{\pm}(t,\vx)&=&\frac{N\sqrt{q}}{2{\kappa^{\hbox{\tiny{(5)}}}}^2}\left[w_0\left(K^{ij}K_{ij}-K^2+\sigma_{\pm}\rt\right)\right.\nonumber\\
&+&\frac{3|\beta|}{8}\rt^2-|\beta|\rt^{ij}\rt_{ij}+2w_0^2\tilde{\Lambda}_{\pm}\bigg]\,,
\eea
where
\bml
\label{7-15}
\bea
\sigma_{\pm}&\equiv&\frac{1}{4} \left(11\pm\sqrt{30}\right)>0\,,\label{7-15a}\\
\tilde{\Lambda}_{\pm}&\equiv&\frac{3}{16|\beta|} \left(13\pm2 \sqrt{30}\right)>0\,.
\eea
\eml
One must remember that so far we have calculated the effective Lagrangean for $g_{\mu\nu}$. But what we really measure in our universe comes from the metric $G_{\mu\nu}=wg_{\mu\nu}$. By considering again that $\tilde{N}=\sqrt{w}N$, $\tilde{N}^i=w\tilde{N}^i$, $\tilde{q}_{ij}=w q_{ij}\Rightarrow \tilde{q}=w^{3/2}q$ and that $\tilde{K}_{ij}=\sqrt{w}K_{ij}$ as one can see from (\ref{4-3c}), the effective Lagrangian is
\bea
\label{7-141}\mathcal{L}_G^{\pm}(t,\vx)&=&\frac{\tilde{N}\sqrt{\tilde{q}}}{2{\kappa^{\hbox{\tiny{(5)}}}}^2}\left(\tilde{K}^{ij}\tilde{K}_{ij}
-\tilde{K}^2+\sigma_{\pm}\rtt\right.\nonumber\\
&+&\frac{3|\beta|}{8}\rtt^2-|\beta|\rtt^{ij}\rtt_{ij}+2\tilde{\Lambda}_{\pm}\bigg)\,.
\eea
Let us define
\bml
\label{7-16}
\bea
\tilde{N}^{\pm}&\equiv&\sqrt{\sigma_{\pm}}\tilde{N}\,,\label{7-16a}\\
{\kappa^{\hbox{\tiny{(4)}}}_{\pm}}^2&\equiv&\frac{{\kappa^{\hbox{\tiny{(5)}}}}^2}{\sqrt{\sigma_{\pm}}}\,,\label{7-16b}\\
\beta_{\pm}&\equiv&\frac{|\beta|}{\sigma_{\pm}}>0\,,\label{7-16c}\\
\Lambda_{\pm}&\equiv&\frac{\tilde{\Lambda}_{\pm}}{\sigma_{\pm}}=\frac{3}{16|\beta|\sigma_{\pm}} \left(13\pm2 \sqrt{30}\right)>0\,,\label{7-16d}
\eea
\eml
in terms of what the resulting Lagrangian is, then,
\bea
\label{7-17}
\mathcal{L}_G^{\pm}(t,\vx)&=&\frac{\tilde{N}^{\pm}\sqrt{\tilde{q}}}{2{\kappa^{\hbox{\tiny{(4)}}}_{\pm}}^2}\bigg(\tilde{K}^{ij}\tilde{K}_{ij}-\tilde{K}^2
+\rtt\nonumber\\
&+&\left.\frac{3\beta_{\pm}}{8}\rtt^2-\beta_{\pm}\rtt^{ij}\rtt_{ij}+2\Lambda_{\pm}\right)\,.
\eea

One important result from the above procedure  is that we have obtained a cosmological constant on the brane without previously introducing it. The cosmological constant obtained in (\ref{7-16d}) has appeared as a result of the Ho\v rava bulk into the brane. Note that this brane cosmological constant is positive, in agreement with a de Sitter-like universe. Moreover, one may verify, by comparing the equations (2.1) and (2.24b) in ref.~\cite{Bemfica7} with equation (\ref{7-17}), that the dispersion relation for the graviton in the linearized version of (\ref{7-17}) is
\be
\label{7-18}
{k^0}^2_{\pm}=\vk^2+\beta_{\pm}\vk^4+2\Lambda_{\pm}\,.
\ee
It is important that $\beta_{\pm}>0$ ($\beta<0$) to avoid a limitation in the graviton energy, otherwise one will get imaginary values for the energy.

\section{Consistency conditions}
\label{sec8}
After the previous considerations, it would be interesting to verify whether negative brane tension is necessary in the compactification scheme. Note that the identity
\be
\label{8-1}
\partial_y\left(w^{\xi}\partial_y w\right)=w^{\xi+1}\left[\xi \frac{\left(\partial_y w\right)^2}{w^2}+\frac{\partial_y^2w}{w}\right]
\ee
enable us to obtain, for an arbitrary $\xi$, the consistency equations
\be
\label{8-2}
\oint dy w^{\xi+1}\left[\left(\xi+\frac{1}{2}\right)\left(\frac{R}{w}-\rf^\mu_\mu\right)+(\xi-1)\rf_{44}\right]=0,
\ee
by considering equations (\ref{4-2}) and by setting $\rf^\mu_\mu\equiv \tilde{\tilde{g}}^{\mu\nu}\rf_{\mu\nu}$. We emphasize that the symbol $\oint$ stands for a integration along the extra dimension $y$ which, by means of the orbifold symmetry, is compact without boundary in such a way that $\oint dy\partial_y\left(w^{\xi}\partial_y w\right)=0$. These consistency equations above may be modified as well by considering
\be
\label{8-3}
\rf^\mu_\mu=-\frac{1}{3}\tilde{\Lambda}^\mu_{\mu}-\frac{4}{3}\tilde{\Lambda}_{44}-\frac{1}{3}{\kappa^{\tin{(5)}}}^2S^\mu_\mu
-\frac{4}{3}{\kappa^{\tin{(5)}}}^2S_{44}
\ee
and
\be
\label{8-4}
\rf_{44}=\frac{2}{3}\tilde{\Lambda}_{44}-\frac{1}{3}\tilde{\Lambda}^\mu_{\mu}-\frac{1}{3}{\kappa^{\tin{(5)}}}^2S^\mu_\mu
+\frac{2}{3}{\kappa^{\tin{(5)}}}^2S_{44}\,,
\ee
where $S=G^{AB}S_{AB}$, as one can see from (\ref{5-1}). The desired form for the consistency equations are, then,
\be
\label{8-5}
\oint dy w^{\xi+1}\left[\left(\xi+\frac{1}{2}\right)\frac{R}{w}+\frac{1}{2}{\kappa^{\tin{(5)}}}^2\,S+\frac{1}{2}\tilde{\Lambda}
+\left(2\xi-\frac{1}{2}\right)\left(\tilde{\Lambda}_{44}+{\kappa^{\tin{(5)}}}^2S_{44}\right)\right]=0\,.
\ee
In the above equation, the definitions $\tilde{\Lambda}\equiv \tilde{\Lambda}^A_A$ and $S\equiv S^A_A$ were used. Notice that the last equation must be valid for any point $t,\vx$ in the (3+1)-dimensional spacetime. Since the space is not full of matter, we choose a point of vacuum (${T_{\tin{B}}}_{AB}=0=T^{(r)}_{\mu\nu}$) so that (\ref{8-5}) must be written as
\be
\label{8-6}
2{\kappa^{\tin{(5)}}}^2\sum_r \tau^{(r)}\omega_r^{\xi+1}= \left(\xi+\frac{1}{2}\right)R\oint dy\, w^\xi+\oint dy\,w^{\xi+1}\left[\frac{1}{2}\tilde{\Lambda}
+\left(2\xi-\frac{1}{2}\right)\tilde{\Lambda}_{44}\right]\,.
\ee
If one chooses $\alpha=\beta=0$, taking $\xi>-1/2$ and remembering that $w>0$, the sum over the tensions seems to be positive. However, since $R$ does not depend on $y$, we can consider our universe where $R$ is practically zero (about $10^{-120}\, M_{P}^{2}$ ~\cite{GKL}). This, in turns, lead us to the necessity of having at least one brane with negative tension. Notice that this problem has disappeared due to the presence of extra terms coming from the higher order terms in 4-spatial curvature in the Ho\v rava-like model we proposed. Of course, it will depend mostly on the choice of the signs of $\alpha$ and $\beta$. In order to fix them, for simplicity we will evaluate the consistency condition for $\xi=1/4$, namely,
\be
\label{8-7}
4{\kappa^{\tin{(5)}}}^2\sum_r \tau^{(r)}\omega_r^{\frac{5}{4}}= \oint dy\,w^{\frac{5}{4}}\tilde{\Lambda}\,.
\ee
The calculation of $\tilde{\Lambda}$ is rather intricate and the details are in appendix \ref{apendice1}. Here we just quote the result
\bea
\label{8-8}
\tilde{\Lambda}&=&-\frac{2}{w^{3/4}}\frac{q^{ik}q^{jl}}{N}D_iD_j\left\{N\left[\beta \rt_{kl}+\left(3\alpha+\frac{\beta}{2}\right)q_{kl}\rt \right]\right\}\nonumber\\
&+&\left[\left(27\alpha+7\beta\right)\frac{(\partial_y w)^2}{w^{7/4}}+4\left(9\alpha+2\beta\right)\partial_y\left(\frac{\partial_y w}{w^{3/4}}\right)\right]\frac{q^{ij}}{2N}D_i\partial_jN+\frac{\beta \rt^{ij}\rt_{ij}+\alpha\rt^2}{2\,w^{3/4}}\nonumber\\
&+&\frac{\beta\,\rt}{2}\left[\frac{(\partial_y w)^2}{w^{7/4}}+2\partial_y\left(\frac{\partial_y w}{w^{3/4}}\right)\right]
-\beta\left[\frac{5}{8}\frac{(\partial_y w)^4}{w^{11/4}}+\partial_y\left(\frac{(\partial_yw)^3}{w^{7/4}}\right)\right]\nonumber\\
&-&3(3\alpha+\beta)\left[\frac{3}{2}\frac{(\partial_y^2w)^2}{w^{3/4}}
-\frac{3}{4}\frac{(\partial_yw)^2\partial_y^2w}{w^{7/4}}-\partial_y\left(\frac{\partial_yw\partial_y^2w}{w^{3/4}}\right)
-2w^{1/4}\partial_y^4w\right]\,.
\eea
So far we left $\alpha$ generic again. But notice that $\tilde{\Lambda}$ contains problematic terms like the square of $\partial^2_yw$, $\partial^4_yw$ and so on. However, all those terms may be eliminated again by imposing the detailed balance condition (\ref{6-7}). In other words, by again fixing $\alpha$ through the equation $3\alpha+\beta=0$. Nothing prevent us from choosing this option, what simplifies last equation as
\bea
\label{8-9}
\frac{16{\kappa^{\tin{(5)}}}^2}{3}\sum_r \tau^{(r)}\omega_r^{\frac{5}{4}}&=&
\beta\left\{\frac{1}{2}\left[\rt^{ij}\rt_{ij}-\frac{1}{3}\rt^2-\frac{4}{N}\left(\rt^{ij}+\frac{1}{2}q^{ij}\rt \right)\right]D_i\partial_j N\,I_1\right.\nonumber\\
&&\phantom{\beta\bigg\{}+\left.\left(\frac{\rt}{2}-\frac{q^{ij}D_i\partial_jN}{N}\right)I_2-\frac{5}{8}I_3\right\}\,.
\eea
Above, we defined the integrals
\be
\label{8-10}
I_1\equiv \oint \frac{dy}{w^{3/4}}>0\,,
\ee
\be
\label{8-11}
I_2\equiv \oint dy \frac{(\partial_y w)^2}{w^{7/4}}>0\,,
\ee
and
\be
\label{8-12}
I_3\equiv \oint dy\frac{(\partial_y w)^4}{w^{11/4}}>0\,.
\ee
The equation (\ref{8-9}) is valid for every point in the (3+1)-spacetime $\Sigma$. To study the sign of $\beta$ we take large radius ($r\to\infty$) in $\Sigma$ where $\rt_{ij}=0$ and $N\to 1$.\cite{footnote04} In this region
\be
\label{8-13}
\frac{16{\kappa^{\tin{(5)}}}^2}{3}\sum_r \tau^{(r)}\omega_r^{\frac{5}{4}}=
-\frac{5\beta}{8}I_3\,.
\ee
Since $\tau^{(r)}$ is a constant, this result must be independent of point chosen to evaluate it. To eliminate negative tension in the branes clearly we have to set $\beta=-|\beta|$, being in accordance with the results obtained in the previous section.

\section{Summary and conclusions}
\label{sec9}

In this work we dealt with a (4+1)-dimensional Ho\v rava-like model treated by means of braneworld models. We have calculated the equation of motion of the model and, by defining a warped metric characterized by the warp factor $w(y)>0$, we showed that warped geometry requires $\lambda=1$. This is interesting as $\lambda \not =1$ has introduced several illness in the original Ho\v rava theory such as ghosts~\cite{Bemfica7}. Furthermore, in the context of brane theory, where $\partial_y w$ is discontinuous across the branes, the extra terms labeled by $\alpha$ and $\beta$ introduces inconsistencies proportional to $(\partial_y^2 w)^2\sim (\delta(y))^2$. However, those inconsistencies are cancelled simply by setting $\alpha=-\beta/3$, what we verified  corresponds exactly to demand detailed balance condition for the quadratic terms in the 4-curvature $\rf$.

The discontinuity of $\partial_y w$ also led us to the Israel's junction condition which, in turns, enabled us to obtain an effective Lagrangian on the brane. Nonetheless, although the resulting Lagrangian corresponded to a (3+1)-dimensional Ho\v rava-like model with an emerging positive cosmological constant, it failed to obey detailed balance condition. By imposing then the detailed balance condition in the quadratic terms in the 3-curvature $\rt$ for this effective theory, positive brane tension demanded $\beta<0$. The most interesting aspect of this result is the fact that this sign for $\beta$ was exactly the expected one for a graviton dispersion relation that have only real energy $k^{0}$.

Despite coincidences among detailed balance condition, positive tension and desired sign for $\beta$, it was important to verify whether the theory permits only positive tension for all branes in the compactification scheme. The response to this task was obtained by studying the consistency conditions in the last section. Again, the detailed balance condition played an essential role by eliminating undesired terms like $(\partial_y^2 w)^2$. Moreover, the answer relied in the sign of $\beta$ that was shown to be negative, in accordance with the results above commented.

The mechanism of detailed balance condition has served as a basis not only for avoiding spreading of constants in the theory, which was one of the arguments used by Ho\v rava, as well as it was essential for the consistency of the model. It has not only guaranteed the improvement of the propagator~\cite{Bemfica6,Bemfica7}, but also eliminated inconsistencies in the present work. Furthermore, it worked in agreement with the expected sign of $\beta$, sometimes imposing it. In fact, we are not aware of the complete implications of the detailed balance condition and its relation in particular with the quadratic term in curvature.\cite{footnote05} So far we have just analyzed its consequences. The present results points to the necessity of a deeper understanding of this mechanism in more general grounds.

\begin{acknowledgments}
This work was partially supported by Funda\c c\~ao de Amparo \`a Pesquisa do Estado de S\~ao Paulo (FAPESP) and Conselho Nacional de Pesquisas (CNPq).
\end{acknowledgments}

\appendix

\section{Evaluation of \boldmath $\tilde{\Lambda}$}
\label{apendice1}

From (\ref{5-2}), it is straightforward to obtain
\be
\label{a1}
\tilde{\Lambda}=\frac{1}{2}\left(\alpha\,\rf^2+\beta\,\rf^{ab}\rf_{ab}\right)+\frac{\tilde{q}^{ab}V_{ab}}{\tilde{N}}\,,
\ee
where the first term in the right hand side of the above equation can be found in (\ref{5-5}). The more intricate term is $\tilde{q}^{ab}V_{ab}$. By defining $v_{ab}\equiv \tilde{N}f_{ab}$ ($v\equiv \tilde{g}^{ab}v_{ab}$) and by keeping in mind that $v_{4i}=0$, it can be expressed as [see (\ref{17})]
\bea
\label{a2}
\tilde{q}^{ab}V_{ab}&=&-\frac{1}{2}\tilde{q}^{ab}\tilde{D}_a\tilde{D}_bv-\tilde{q}^{ac}\tilde{q}^{bd}\tilde{D}_a\tilde{D}_bv_{cd}\nonumber\\
&=&-\frac{1}{2}\partial_y^2 v-\frac{1}{2}\tilde{q}^{ij}\tilde{D}_i\tilde{D}_jv-\partial_y^2v_{44}
-\tilde{q}^{ij}\tilde{D}_4\tilde{D}_iv_{4j}-\tilde{q}^{ij}\tilde{D}_i\tilde{D}_4v_{4j}-\tilde{q}^{ik}\tilde{q}^{jl}\tilde{D}_i\tilde{D}_jv_{kl}\,.
\nonumber\\
\eea
We are benefited of the identity
\be
\label{a3}
\gf^4_{4a}=\gf^a_{44}=0\,.
\ee
Note that
\be
\label{a4}
\gf^4_{ij}=-\frac{\tilde{g}_{ij}}{2}\frac{\partial_y w}{w}\,,
\ee
while
\be
\label{a5}
\gf^i_{4j}=\frac{\delta^i_j}{2}\frac{\partial_y w}{w}\,.
\ee
Pure spatial indexes lead us to
\be
\label{a6}
\gf^i_{jk}=\frac{q^{il}}{2}\left(2\partial_{(j}q_{k)l}-\partial_lq_{jk}\right)=\gt^i_{jk}\,.
\ee
Although $v_{i4}=0$, this is not true for $\tilde{D}_av_{4i}$. The important relations one needs at this point are
\be
\label{a7}
\tilde{D}_iv_{4j}=\frac{1}{2}\frac{\partial_y w}{w}\left(\tilde{q}_{ij}v_{44}-v_{ij}\right)\,,
\ee
\be
\label{a8}
\tilde{D}_4v_{ij}=w\partial_y\left(\frac{v_{ij}}{w}\right)\,,
\ee
\bea
\label{a81}
\tilde{D}_iv_{jk}=D_iv_{jk}\,,
\eea
together with $\tilde{D}_4v_{i4}=0$. With those relations in mind one is able to get
\bea
\label{a9}
\tilde{q}^{ij}\tilde{D}_i\tilde{D}_jv&=&\tilde{q}^{ij}\left(\partial_i\partial_jv-\gf^{a}_{ij}\partial_av\right)\nonumber\\
&=&\frac{q^{ij}}{w}D_i\partial_jv+\frac{3}{2}\frac{\partial_yw\partial_yv}{w}\,,
\eea
\bea
\label{a10}
\tilde{q}^{ij}\tilde{D}_4\tilde{D}_iv_{4j}&=&\tilde{q}^{ij}\left(\partial_y\tilde{D}_iv_{4j}
-\gf^a_{4i}\tilde{D}_av_{4j}-\gf^a_{4j}\tilde{D}_iv_{4a}\right)\nonumber\\
&=&\frac{1}{2}\partial_y\left[\frac{\partial_y w\left(4v_{44}-v\right)}{w}\right]\,,
\eea
\bea
\label{a11}
\tilde{q}^{ij}\tilde{D}_i\tilde{D}_4v_{4j}&=&\tilde{q}^{ij}\left(
-\gf^l_{i4}\tilde{D}_lv_{4j}-\gf^l_{i4}\tilde{D}_4v_{lj}-\gf^4_{ij}\partial_yv_{44}\right)\nonumber\\
&=&\frac{1}{2}\frac{\partial_yw}{\sqrt{w}}\left(\frac{4v_{44}-v}{\sqrt{w}}\right)\,,
\eea
and
\bea
\label{a12}
\tilde{q}^{ik}\tilde{q}^{jl}\tilde{D}_i\tilde{D}_jv_{kl}&=&\tilde{q}^{ik}\tilde{q}^{jl}
\left(\partial_i\tilde{D}_jv_{kl}-\gf^a_{ij}\tilde{D}_av_{kl}-\gf^a_{ik}\tilde{D}_jv_{al}-\gf^a_{il}\tilde{D}_jv_{ka}\right)\nonumber\\
&=&\frac{q^{ik}q^{jl}}{w^2}D_iD_jv_{kl}+\frac{1}{2}\frac{\partial_yw\partial_y\left(v-v_{44}\right)}{w}+\frac{(\partial_yw)^2}{w^2}(4v_{44}-v)\,.
\eea
By collecting the results (\ref{a7})--(\ref{a12}) into (\ref{a2}), dividing it by $\tilde{N}=\sqrt{w}N$, writing $v_{ab}=\sqrt{w}Nf_{ab}$ and then inserting it into (\ref{a1}) one finally gets the result quoted in (\ref{8-8}).


\begin{thebibliography}{21}
\expandafter\ifx\csname natexlab\endcsname\relax\def\natexlab#1{#1}\fi
\expandafter\ifx\csname bibnamefont\endcsname\relax
  \def\bibnamefont#1{#1}\fi
\expandafter\ifx\csname bibfnamefont\endcsname\relax
  \def\bibfnamefont#1{#1}\fi
\expandafter\ifx\csname citenamefont\endcsname\relax
  \def\citenamefont#1{#1}\fi
\expandafter\ifx\csname url\endcsname\relax
  \def\url#1{\texttt{#1}}\fi
\expandafter\ifx\csname urlprefix\endcsname\relax\def\urlprefix{URL }\fi
\providecommand{\bibinfo}[2]{#2}
\providecommand{\eprint}[2][]{\url{#2}}

\bibitem[{\citenamefont{Stelle}(1977)}]{Stelle}
\bibinfo{author}{\bibfnamefont{K.~S.} \bibnamefont{Stelle}},
  \bibinfo{journal}{Phys. Rev. D} \textbf{\bibinfo{volume}{16}},
  \bibinfo{pages}{953} (\bibinfo{year}{1977}).

\bibitem[{\citenamefont{Horava}(2009)}]{Horava2}
\bibinfo{author}{\bibfnamefont{P.}~\bibnamefont{Horava}},
  \bibinfo{journal}{Phys. Rev. D} \textbf{\bibinfo{volume}{79}},
  \bibinfo{pages}{084008} (\bibinfo{year}{2009}), \eprint{0901.3775}.

\bibitem[{foo({\natexlab{a}})}]{footnote01}
\bibinfo{note}{We will return to this point in a latter time and define
  detailed balance condition.}

\bibitem[{\citenamefont{Orlando and Reffert}(2009)}]{Orlando}
\bibinfo{author}{\bibfnamefont{D.}~\bibnamefont{Orlando}} \bibnamefont{and}
  \bibinfo{author}{\bibfnamefont{S.}~\bibnamefont{Reffert}},
  \bibinfo{journal}{Class.Quant.Grav.} \textbf{\bibinfo{volume}{26}},
  \bibinfo{pages}{155021} (\bibinfo{year}{2009}), \eprint{0905.0301}.

\bibitem[{\citenamefont{Bemfica and Gomes}(2011{\natexlab{a}})}]{Bemfica6}
\bibinfo{author}{\bibfnamefont{F. S.}~\bibnamefont{Bemfica}} \bibnamefont{and}
  \bibinfo{author}{\bibfnamefont{M.}~\bibnamefont{Gomes}},
  \bibinfo{journal}{Phys.Rev.} \textbf{\bibinfo{volume}{D84}},
  \bibinfo{pages}{084022} (\bibinfo{year}{2011}{\natexlab{a}}),
  \eprint{1108.5979}.

\bibitem[{\citenamefont{Bemfica and Gomes}(2011{\natexlab{b}})}]{Bemfica7}
\bibinfo{author}{\bibfnamefont{F. S.}~\bibnamefont{Bemfica}} \bibnamefont{and}
  \bibinfo{author}{\bibfnamefont{M.}~\bibnamefont{Gomes}}
  (\bibinfo{year}{2011}{\natexlab{b}}), \eprint{1111.5779}.

\bibitem[{\citenamefont{Randall and Sundrum}(1999)}]{RS}
\bibinfo{author}{\bibfnamefont{L.}~\bibnamefont{Randall}} \bibnamefont{and}
  \bibinfo{author}{\bibfnamefont{R.}~\bibnamefont{Sundrum}},
  \bibinfo{journal}{Phys.Rev.Lett.} \textbf{\bibinfo{volume}{83}},
  \bibinfo{pages}{4690} (\bibinfo{year}{1999}), \eprint{hep-th/9906064}.

\bibitem[{\citenamefont{Abdalla et~al.}(2010)\citenamefont{Abdalla, Guimaraes,
  and Hoff~da Silva}}]{AGJ}
\bibinfo{author}{\bibfnamefont{M.}~\bibnamefont{Abdalla}},
  \bibinfo{author}{\bibfnamefont{M.}~\bibnamefont{Guimaraes}},
  \bibnamefont{and} \bibinfo{author}{\bibfnamefont{J. M.}~\bibnamefont{Hoff~da
  Silva}}, \bibinfo{journal}{JHEP} \textbf{\bibinfo{volume}{1009}},
  \bibinfo{pages}{051} (\bibinfo{year}{2010}), \eprint{1001.1075}.

\bibitem[{\citenamefont{Hoff~da Silva and Dias}(2011)}]{ehNOIS}
\bibinfo{author}{\bibfnamefont{J. M.}~\bibnamefont{Hoff~da Silva}}
  \bibnamefont{and} \bibinfo{author}{\bibfnamefont{M.}~\bibnamefont{Dias}},
  \bibinfo{journal}{Phys.Rev.} \textbf{\bibinfo{volume}{D84}},
  \bibinfo{pages}{066011} (\bibinfo{year}{2011}), \eprint{1107.2017}.

\bibitem[{\citenamefont{Gibbons et~al.}(2001)\citenamefont{Gibbons, Kallosh,
  and Linde}}]{GKL}
\bibinfo{author}{\bibfnamefont{G.~W.} \bibnamefont{Gibbons}},
  \bibinfo{author}{\bibfnamefont{R.}~\bibnamefont{Kallosh}}, \bibnamefont{and}
  \bibinfo{author}{\bibfnamefont{A.~D.} \bibnamefont{Linde}},
  \bibinfo{journal}{JHEP} \textbf{\bibinfo{volume}{0101}}, \bibinfo{pages}{022}
  (\bibinfo{year}{2001}), \eprint{hep-th/0011225}.

\bibitem[{\citenamefont{Leblond et~al.}(2001)\citenamefont{Leblond, Myers, and
  Winters}}]{FR}
\bibinfo{author}{\bibfnamefont{F.}~\bibnamefont{Leblond}},
  \bibinfo{author}{\bibfnamefont{R.~C.} \bibnamefont{Myers}}, \bibnamefont{and}
  \bibinfo{author}{\bibfnamefont{D.~J.} \bibnamefont{Winters}},
  \bibinfo{journal}{JHEP} \textbf{\bibinfo{volume}{0107}}, \bibinfo{pages}{031}
  (\bibinfo{year}{2001}), \eprint{hep-th/0106140}.

\bibitem[{\citenamefont{Goldberger and Wise}(1999)}]{GW}
\bibinfo{author}{\bibfnamefont{W.~D.} \bibnamefont{Goldberger}}
  \bibnamefont{and} \bibinfo{author}{\bibfnamefont{M.~B.} \bibnamefont{Wise}},
  \bibinfo{journal}{Phys.Rev.Lett.} \textbf{\bibinfo{volume}{83}},
  \bibinfo{pages}{4922} (\bibinfo{year}{1999}), \eprint{hep-ph/9907447}.

\bibitem[{\citenamefont{Henneaux et~al.}(2010)\citenamefont{Henneaux,
  Kleinschmidt, and Gomez}}]{Henneaux}
\bibinfo{author}{\bibfnamefont{M.}~\bibnamefont{Henneaux}},
  \bibinfo{author}{\bibfnamefont{A.}~\bibnamefont{Kleinschmidt}},
  \bibnamefont{and} \bibinfo{author}{\bibfnamefont{G.~L.} \bibnamefont{Gomez}},
  \bibinfo{journal}{Phys. Rev. D} \textbf{\bibinfo{volume}{81}},
  \bibinfo{pages}{064002} (\bibinfo{year}{2010}), \eprint{0912.0399}.

\bibitem[{\citenamefont{Blas et~al.}(2009)\citenamefont{Blas, Pujolas, and
  Sibiryakov}}]{Blas}
\bibinfo{author}{\bibfnamefont{D.}~\bibnamefont{Blas}},
  \bibinfo{author}{\bibfnamefont{O.}~\bibnamefont{Pujolas}}, \bibnamefont{and}
  \bibinfo{author}{\bibfnamefont{S.}~\bibnamefont{Sibiryakov}},
  \bibinfo{journal}{JHEP} \textbf{\bibinfo{volume}{10}}, \bibinfo{pages}{029}
  (\bibinfo{year}{2009}), \eprint{0906.3046}.

\bibitem[{foo({\natexlab{b}})}]{footnote02}
\bibinfo{note}{For technical details on those calculations see, for instance,
  refs.~\cite{Wald,Thiemann}.}

\bibitem[{\citenamefont{Wald}(1984)}]{Wald}
\bibinfo{author}{\bibfnamefont{R.~M.} \bibnamefont{Wald}},
  \emph{\bibinfo{title}{General Relativity}} (\bibinfo{publisher}{The
  University of Chicaco Press}, \bibinfo{address}{Chicaco and London},
  \bibinfo{year}{1984}).

\bibitem[{\citenamefont{Israel}(1966)}]{Israel}
\bibinfo{author}{\bibfnamefont{W.}~\bibnamefont{Israel}}, \bibinfo{journal}{Il
  Nuovo Cimento B} \textbf{\bibinfo{volume}{44}}, \bibinfo{pages}{1}
  (\bibinfo{year}{1966}).

\bibitem[{foo({\natexlab{c}})}]{footnote03}
\bibinfo{note}{Remember we are considering $Z_2$ symmetry in the $y$
  direction.}

\bibitem[{foo({\natexlab{d}})}]{footnote04}
\bibinfo{note}{Obviously, it is in accordance with our previous vacuum case
  particularization.}

\bibitem[{foo({\natexlab{e}})}]{footnote05}
\bibinfo{note}{It is worth saying that, at least regarding ultraviolet behavior
  of the propagator, there is a necessity of the detailed balance condition
  only in the quadratic terms in curvature, a soft detailed balance condition,
  despite the presence of higher order terms in the action, as shown in
  ref.~\cite{Bemfica7}.}

\bibitem[{\citenamefont{Thiemann}(2007)}]{Thiemann}
\bibinfo{author}{\bibfnamefont{T.}~\bibnamefont{Thiemann}},
  \emph{\bibinfo{title}{Modern canonical quantum general relativity}}
  (\bibinfo{publisher}{Cambridge University Press}, \bibinfo{year}{2007}).

\end{thebibliography}

\end{document}